\documentclass[twocolumn,aps,amsmath,amssymb,showpacs,pre]{revtex4}
\usepackage{graphicx}
\usepackage{epsfig}
\usepackage{amsmath}
\bibstyle{apsrev.bib}

\begin{document}
\input epsf.sty

\title{Competition between quenched disorder and long-range connections: A
  numerical study of diffusion}

\author{R\'obert Juh\'asz}
 \email{juhasz@szfki.hu} 
\affiliation{Research Institute for Solid
State Physics and Optics, H-1525 Budapest, P.O.Box 49, Hungary}

\date{\today}

\begin{abstract}
The problem of random walk is considered in one dimension 
in the simultaneous presence of a 
quenched random force field and long-range connections the probability of
which decays with the distance algebraically as $p_l\simeq \beta l^{-s}$. 
The dynamics are studied mainly by a numerical 
strong disorder renormalization group method.  
According to the results, for $s>2$ the long-range connections are irrelevant
and the mean-square displacement increases as 
$\langle x^2(t)\rangle\sim (\ln t)^{2/\psi}$ with the barrier
exponent $\psi=1/2$, which is known in one-dimensional random environments. 
For $s<2$, instead, the quenched disorder is found to be 
irrelevant and 
the dynamical exponent is $z=1$ like in a homogeneous environment. 
At the critical point, $s=2$, the interplay between 
quenched disorder and long-range
connections results in activated scaling, however, with a non-trivial barrier
exponent $\psi(\beta)$, which decays continuously with $\beta$
but is independent of the form of the quenched disorder.
Upper and lower bounds on $\psi(\beta)$ are established and numerical
estimates are given for various values of $\beta$. 
Beside random walks, accurate numerical estimates of the
graph dimension and the resistance exponent are given for various 
values of $\beta$ at $s=2$.    
\end{abstract}


\maketitle

\newcommand{\bc}{\begin{center}}
\newcommand{\ec}{\end{center}}
\newcommand{\be}{\begin{equation}}
\newcommand{\ee}{\end{equation}}
\newcommand{\beqn}{\begin{eqnarray}}
\newcommand{\eeqn}{\end{eqnarray}}

\vskip 2cm
\section{Introduction}

Long-range interactions are known to effect the dynamics of various systems
such as equilibrium spin models \cite{lb}, non-equilibrium processes 
\cite{Hbook} or diffusion \cite{bouchaud}. 
A well-known generalization of normal diffusion is the L\'evy flight 
\cite{weierstrass,bouchaud}, where 
long jumps are allowed with probabilities decaying
algebraically with the distance as $p_l\sim l^{-1-f}$. 
These long jumps lead to superdiffusion with
an anomalous dynamical exponent $z=f$ for $f<2$, while for fast enough decay ($f>2$) the diffusion is normal ($z=2$). 
Contrary to long-range interactions, local heterogeneities, frequently slow down the dynamics, which manifests itself in an increased dynamical exponent compared to that of the pure system. 
This is also the case for diffusion in various types of random environments
\cite{havlin,bouchaud}. Here, the most dramatic slowing down is induced by the random-force
disorder, where the direction of the local force acting on the particle is
random. 
This type of disorder is known to be relevant if the dimension $d$ is less than $2$ \cite{luck,fisher}. 
In one dimension, the mean square displacement increases in time 
as $\overline{\langle x^2(t)\rangle}\sim (\ln t)^4$ \cite{sinai}, where
$\langle\cdot\rangle$ denotes average over stochastic histories for a given
random environment and the overbar denotes average over random environments. 
This type of ultra slow dynamics corresponds formally to an infinite 
dynamical exponent.      

It is an intriguing question what happens when both long-range jumps and
random force-disorder are present in a system and whether there is a
non-trivial dynamical behavior due to the interplay between them.
In case of diffusion, a 
dynamic renormalization group analysis showed that the dynamical exponent
equals to the L\'evy index $f$ even in the presence of 
weak disorder for $f<2$ \cite{fogedby}. Thus occasional long jumps enable the
particle to escape from trapping regions of the random environment and
disorder is irrelevant in the above sense. 

An alternative form of long-range interactions is realized when the interaction
strength is constant but only certain pairs of sites interact
with each other which are selected randomly with a probability 
decaying asymptotically with the distance $l$ of the sites as 
$p_l\simeq \beta l^{-s}$.   
This paradigm arises e.g. in the context of magnetic \cite{cms} and 
conducting \cite{cc} properties of linear polymers with 
crosslinks between certain monomers
or in models of social or communication networks which exhibit the small-world phenomenon \cite{newman,kleinberg}. 
The interactions are in this case represented by a random graph 
which is composed of short edges between neighboring sites 
and long edges between certain remote sites. 
This model carries thus an intrinsic quenched disorder due to the interaction
network as opposed to the former paradigm which can be regarded 
as an annealed counterpart of the latter.  
Here, even the geometrical properties of the interaction graph are
non-trivial \cite{nw,bb,sc,mam,coppersmith,biskup,gbs}. 
Adding long edges to a one-dimensional lattice, the resulting graph is
quasi-one-dimensional if 
$s>2$, whereas it is formally infinite-dimensional if $s<2$ \cite{biskup}.
These domains are separated by a critical point at $s=2$, where 
the graph dimension is conjectured to be finite and to depend
on the prefactor $\beta$ \cite{bb}.     
The geometrical properties of the underlying interaction network indicate  
that such models may show non-conventional dynamical behavior. 
Indeed, the contact process with long-range interactions of this type is
conjectured to exhibit activated dynamical scaling at $s=2$ with
$\beta$-dependent critical exponents \cite{prl,ojcm}. 

One can pose the question of how the dynamics in general 
are affected by the simultaneous presence of this type of long-range connections and quenched local disorder. 
In this paper we shall study this issue by considering a simple dynamical
process, the random walk with random-force disorder on graphs containing
long edges with algebraically decaying probability. 
Most efforts will be devoted to the critical point $s=2$ 
where the graph dimension is finite and which gained less attention
in earlier numerical studies. 
As the dynamics are expected to be slow here, we shall mainly apply a 
numerical strong disorder renormalization method \cite{im} carried out in the
configuration space \cite{jsi,monthus,vulpiani,juhasz} rather than Monte Carlo
simulations. 
Beside the above investigations, the non-trivial geometrical
properties such the graphs dimension and the random-walk dimension will be
estimated at $s=2$, as well.
Our results, which are accurate for moderate $\beta$ at the
critical point $s=2$, indicate that the dynamics are still activated, i.e. 
the mean-square displacement increases as  
$\overline{\langle x^2(t)\rangle}\sim (\ln t)^{2/\psi(\beta)}$, however the barrier exponent
$\psi(\beta)$ differs from $1/2$ and 
decreases continuously with the prefactor $\beta$. 

The rest of the paper is organized as follows. 
The precise definition of the model is given in In Sec. \ref{model}. 
In Sec. \ref{sp}, numerical estimates of the shortest-path dimension are
presented. In Sec. \ref{hrw}, bounds on the random walk dimension are
established, as well as numerical estimates are given. 
In Sec. \ref{rwre}, the strong disorder renormalization group method is 
reviewed and applied to the problem of random walks with long-range connections in random environment. 
Finally, the results are discussed in Sec. \ref{discussion}.

\section{The model}
\label{model}

The finite connectivity networks to be studied are defined as follows. 
Assume that the sites are numbered by the integers $1,2,\dots,N$ and 
define the distance between site $i$ and $j$ as 
$l_{ij}=\min (|i-j|,N-|i-j|)$. In words, the sites are arranged on a ring with unit spacing between them. 
Then all pairs of sites with a distance $l_{ij}=1$ (i.e. neighboring sites on the
ring) are connected with an edge and 
all pairs with $l_{ij}>1$ are connected independently with the
probability \cite{an,bb}
\be 
P(l)=1-\exp (-\beta l^{-s}), 
\label{pl}
\ee
where $\beta$ and $s$ are positive constants. 
For large $l$, this probability has the asymptotic form: 
$p(l)\simeq \beta l^{-s}$.

On each realization of the above connectivity network a
continuous time random walk is considered, which is specified 
by the set of i.i.d quenched random transition rates $w_{ij}$ from site $i$ to site $j$.
Note that the transition rates through long connections ($l_{ij}>1$) 
are random just as those through short connections ($l_{ij}=1$).   
Note, furthermore, that the source of randomness is twofold in this model. 
First, the random topology of the connectivity network, second, the set of random transition rates. 

\section{Shortest paths}
\label{sp}

From the point of view of dynamical processes like random walk, 
the relevant metric is not the distance $l_{ij}$ but  
the length $\ell$ of shortest paths (or chemical distance) between two sites, which is given by the number of (short or long) edges which constitute the path. 
If $s>2$, the expected value of the length of long edges is finite therefore 
the average shortest path length is asymptotically proportional to the
distance $l$; in other words, the system is quasi-one-dimensional. 
At the critical point $s=2$, mean shortest path length has been 
conjectured to grow
algebraically with the distance: 
\be 
\overline{\ell}\sim l^{1/d_g(\beta)},  
\label{dg}
\ee
where the graph dimension $d_g(\beta)$ depends on $\beta$ \cite{bb}. 
Indeed, later a lower bound on $d_g(\beta)$ has been proven: 
$d_g(\beta)\ge \ln 3/\ln(3-\delta)$, where 
$\delta =1-e^{-\beta/9}$, as well as an upper  bound: 
$d_g(\beta)\le 1/(1-\beta)$ for $\beta<1$ \cite{coppersmith}. 
The latter inequality is related to the presence of so called cut points, which are defined as follows.  
First, instead of the cyclic graphs defined in the previous section, 
let us consider acyclic ones, which are defined in the same
way except that the modified distance $l_{ij}=|i-j|$ is used there. 
In such acyclic graphs, a cut point is a site the removal of which 
(and all of its edges) results in that the graph becomes disconnected.     
The expected number of such points tends to zero for $s<2$, is $O(N)$ for
$s>2$ and $O(N^{1-\beta})$ for $s=2$, $\beta<1$. \cite{an,coppersmith}. 
Obviously, the shortest path between site $1$ and site $N$ must
contain all cut points of the graph, which implies the above lower bound
on $d_g(\beta)$. 
For $s<2$, the graph dimension is formally infinite, for details see 
Refs. \cite{bb,coppersmith,biskup}.  

\begin{figure}[h]
\includegraphics[width=0.9\linewidth]{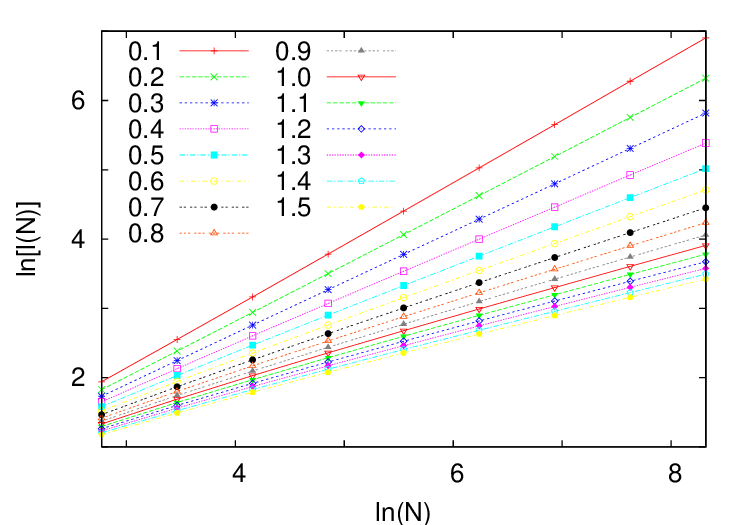}
\caption{\label{fig1} (Color online) The logarithm of the numerically calculated mean length of shortest paths between sites in a
  distance $N/2$ plotted against $\ln N$ for different values $\beta$ at
  $s=2$. The statistical errors are smaller than the size of the symbols. }
\end{figure}

\begin{figure}[h]
\includegraphics[width=0.8\linewidth]{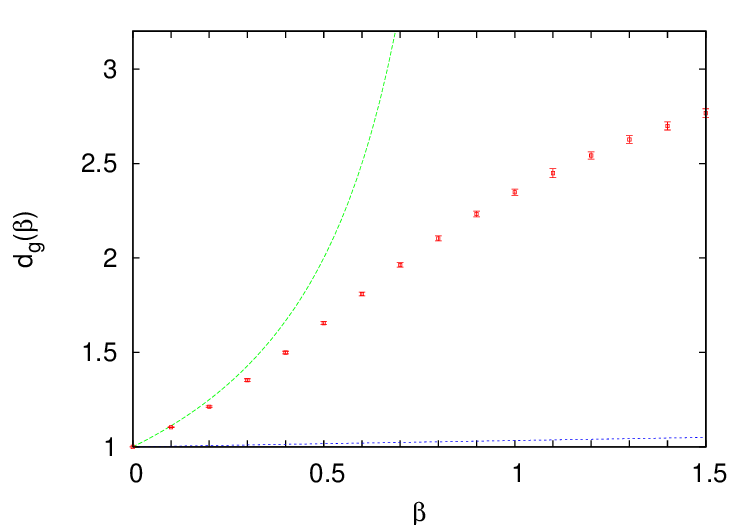}
\caption{\label{fig2} (Color online) Graph dimensions estimated from the data shown in
  Fig. \ref{fig1} for different values of $\beta$. The solid lines represent
  the rigorous bounds on $d_g(\beta)$ \cite{coppersmith}.}
\end{figure}

We considered networks of size $N=2^n$ with $n=4,5,\dots,12$ and
calculated the length of shortest path between pairs of sites in a distance
$l=N/2$ numerically by a breadth-first-search. 
$10^6$ independent networks have been generated for each size and $\ell$ 
between $N/8$ pairs of sites has been calculated in each realization. 
The mean length of shortest paths obtained in this way is plotted 
against $N$ in Fig. \ref{fig1}. 
As can be seen the data are compatible with the power-law in Eq. (\ref{dg}) 
and the estimated graph dimensions, which are obtained extrapolating the
effective two-point exponents to $N\to\infty$, lie between the rigorous bounds
quoted above, see Fig. (\ref{fig2}) and Table \ref{table}.  

\section{Random walk and electrical resistance}
\label{hrw}

Before discussing the random walk with quenched random transition rates, 
we investigate the homogeneous model, where the transition rates are equal, 
say $w_{ij}=1$. For this model, the random walk is known to be recurrent if
$s\ge 2$ and transient if $1<s<2$ \cite{berger,jespersen}.  
On the grounds of calculations in similar but analytically tractable random
networks \cite{juhasz2}, superdiffusion is expected for $s=2$, i.e. the
mean-square-displacement in an infinite network is expected to behave 
asymptotically as 
\be
\overline{\langle x^2(t)\rangle}\sim t^{2/d_w},
\label{dw}
\ee  
where the random-walk dimension $d_w$ \cite{gaa} is less than $2$. 
Although, the exact value of $d_w$ is unknown, we can establish bounds on it in
terms of the graph dimension, then we will give numerical estimates.
In order to do this, we make use of the well-known relationship between
diffusion and resistor networks \cite{havlin,bouchaud}. 
For symmetric random walks (where $w_{ij}=w_{ji}$ holds) on an
arbitrary graph, a resistor network can be defined where the edges of the
graph are resistors with a resistance $R_{ij}=1/w_{ij}$.   
In case of homogeneous transition rates, 
the mean first passage time from one site to another one
whose distance is in the order of the graph diameter 
(e.g. site $1$ and site $N/2+1$ for cyclic graphs) is 
related to the effective resistance $R(N)$ of the corresponding resistor
network between the two sites as $t\sim NR(N)$ for large $N$ \cite{gefen,juhasz2}.
Assuming that Eq. (\ref{dw}) holds, the average effective resistance behaves asymptotically as 
\be
R(N)\sim N^{\zeta},
\label{zeta}
\ee 
where the resistance exponent $\zeta$ is related to 
the random-walk dimension as 
\be 
d_w=1+\zeta.
\label{dwzeta}
\ee
Bounds on the random-walk dimension at $s=2$ can be obtained by simple
considerations as follows. 
In a cyclic graph of size $N$ let us consider two sites in a distance $l=N/2$. 
The length of the shortest path between them is $O(N^{1/d_g})$ according
to Eq. (\ref{dg}). 
From Eq. (\ref{dw}) we obtain, that the mean first passage time 
from one site to the other one is $O(N^{d_w})$.   
Deleting now all edges other than those contained in the shortest path, 
one obtains a one-dimensional system in which the diffusion is normal. So, 
the expected first passage time in this one-dimensional chain is $O(N^{2/d_g})$ 
and it is easy to see that this time must not be greater than that in the original network, 
implying the inequality 
\be
d_w\ge 2/d_g.
\label{lower}
\ee
Consider again the above one-dimensional chain between the two sites. The
effective resistance of this chain is $O(N^{1/d_g})$, and this must be larger
than the effective resistance between the two points in the original network,
which is $O(N^{\zeta})$, since removing links from the original network does not
decrease the effective resistance. So, we have the relation $1/d_g\ge\zeta$
or, using Eq. (\ref{dwzeta}),  
\be
d_w\le 1+1/d_g.
\label{upper}
\ee
Since, for $\beta>0$, $d_g(\beta)>1$ holds, we have $d_w<2$,
i.e. superdiffusion when $\beta>0$.
Furthermore, since both the lower and upper bounds decrease with $\beta$ and for $\beta=0$ the above relations hold as equalities, we conclude that  
the random-walk dimension must also depend on $\beta$. 

\begin{figure}[h]
\includegraphics[width=0.9\linewidth]{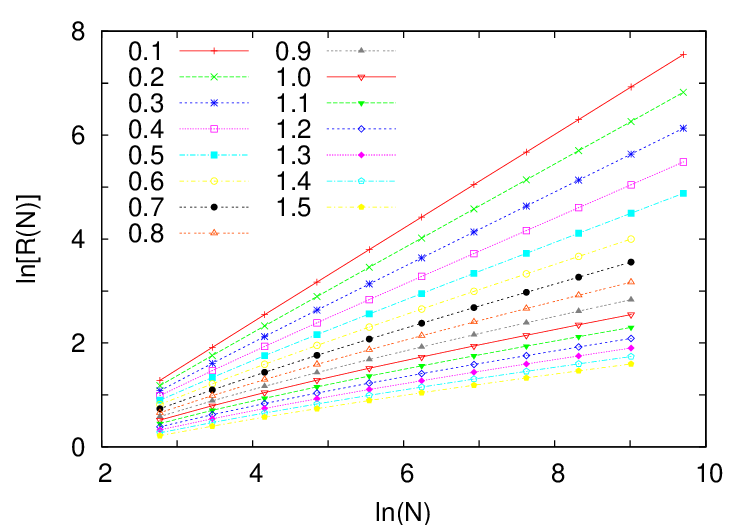}
\caption{\label{fig3} (Color online) The logarithm of the numerically calculated average
  resistance between sites in a
  distance $N/2$ plotted against $\ln N$ for different values $\beta$ at
  $s=2$. The statistical errors are smaller than the size of the symbols. }
\end{figure}

\begin{figure}[h]
\includegraphics[width=0.8\linewidth]{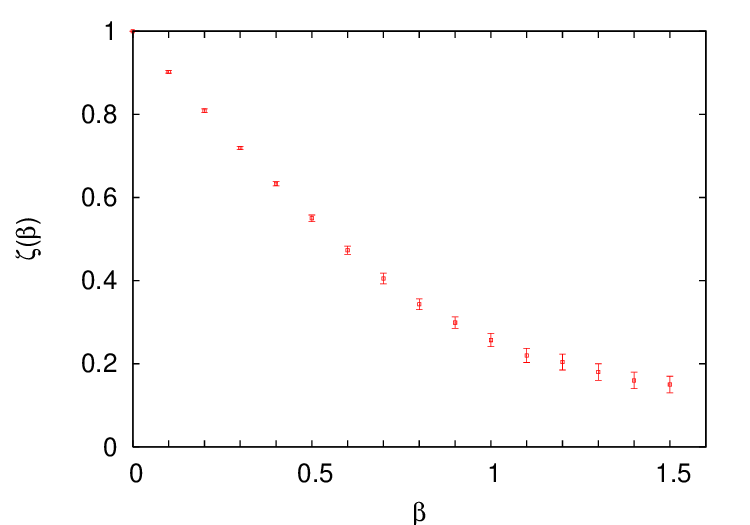}
\caption{\label{fig4} (Color online) Resistance exponents estimated from the data shown in
  Fig. \ref{fig1}.}
\end{figure}

In order to estimate $d_w(\beta)$, we have numerically calculated the
effective resistance between sites $1$ and $N/2+1$ in cyclic 
networks of size $N$ by
the method described in the next section. 
We considered system sizes $N=2^n$ with $n=4,5,\dots,14$ and generated $10^6$
independent networks for $n\le 12$ and $10^5$ for $n=13,14$. 
The average resistance is plotted against $N$ for different values of $\beta$
in Fig. \ref{fig3}. As can be seen, the data are agreement with the power-law
in Eq. (\ref{zeta}) with a $\beta$-dependent resistance exponent. 
The estimated values of $\zeta(\beta)$ obtained by extrapolating 
the two-point effective
exponents are plotted for different values of $\beta$ in Fig. \ref{fig4} 
and are given in Table \ref{table}.
The numerical estimates of $d_w(\beta)$ and $d_g(\beta)$ are found to 
satisfy the inequalities (\ref{lower}) and (\ref{upper}).  

\section{Random walk in random environment}
\label{rwre}

\subsection{Finite graph dimension ($s\ge 2$)}

After having studied the geometry and homogeneous random walks we turn to the
full problem, where the transition rates are quenched random variables. 
In one-dimension, which corresponds to $\beta=0$ in our model, the 
mean-square displacement in an infinite system behaves as 
$\overline{\langle x^2(t)\rangle}\sim (\ln t)^4$ \cite{sinai}. 
It is intuitively obvious, that adding long edges to the one-dimensional
chain results in that the mean-square displacement increases faster with time
since considerably shorter paths between far away sites are opened.  
For $s>2$, on the other hand, the number of cut points, which must be contained
in shortest paths, is $O(N)$. Therefore the logarithm of the typical time to travel through a
domain of length $N$, is at least $O(N^{1/2})$.
As a consequence, for $s>2$, the dynamics must obey the asymptotic law 
\be 
\overline{\langle x^2(t)\rangle}\sim (\ln t)^{2/\psi},
\label{activated}
\ee
with the barrier exponent $\psi=1/2$ as in one dimension, 
and only the proportionality constant is expected to be modified by the
presence of long edges. 

At the critical point $s=2$, $\beta>0$, the form in Eq. (\ref{activated})
with $\psi=1/2$ is a lower bound on $\overline{\langle x^2(t)\rangle}$. 
An upper bound for $\beta<1$ can be easily established since the number of cut
points is $O(N^{1-\beta})$, thus the logarithm of mean first passage time 
must be typically at least $O(N^{(1-\beta)/2})$.
Hence we expect for $s=2$ and at least for $\beta<1$ the activated dynamics
given in Eq. (\ref{activated}) to hold with some barrier exponent
$\psi(\beta)$, which may depend on $\beta$ and lies in the range 
\be
(1-\beta)/2\le \psi(\beta)\le 1/2.
\ee
This type of ultra-slow dynamics suggests that a strong disorder
renormalization group (SDRG) method can be applied to the process. 
This method has been
originally constructed for studying disordered quantum spin chains \cite{mdh},
later it has been adapted for investigating dynamical processes, 
for a review see \cite{im}. 
For random walks in one-dimensional random environments the method has been
formulated in terms of the potential barriers \cite{dmf}. 
Apart from tree-like graphs, however, potential cannot be defined in 
general, and the
method has to be formulated in terms of transition rates 
\cite{jsi,juhasz,monthus,vulpiani}.

In the SDRG procedure, sites with short waiting time are successively
eliminated, and the remaining transition rates are modified such that the long
time dynamical behavior of the system remains unaltered.  
The detailed scheme of the method is the following. 
Let us introduce the exit rate from site
$i$ as $\Omega_i=\sum_jw_{ij}$, where the summation goes over all sites
connected with site $i$. The inverse of $\Omega_i$ gives the expected waiting
time at site $i$: $\tau_i=1/\Omega_i$.   
The elementary step of the procedure is that the site which has the smallest waiting time is chosen and removed 
from the network together with their edges. 
The transition rates $w_{jk}$ between sites which were connected to the
eliminated site $i$, are modified as
\be 
\tilde w_{jk} = w_{jk}+  w_{ji}w_{ik}/\Omega_i.
\label{rg}
\ee
If these sites were not connected before the decimation 
(i.e.  $w_{jk}= w_{kj}=0$) then a new edge is 
created between them with the rates given above.
Through the transition rates, the exit rates and the waiting times of sites
neighboring to the decimated one are also renormalized.  
This decimation step is then iterated starting from a finite system of size
$N$ until only one site remains with an
 effective waiting time $\tau_N=1/\Omega_N$. 
The renormalization rule in Eq. (\ref{rg}) can obtained from the condition 
that the ratios of steady state probabilities of all sites not yet decimated 
(active sites) remain unchanged when a site is eliminated. 
Concerning the dynamics, the replacement of the original system by the one 
site smaller effective one is an approximation in the sense that 
the waiting time in the eliminated site is neglected. 
In certain systems, however, the distribution of the logarithm of waiting
times becomes broader and broader as the ratio of active sites is
decreased. In these systems, which are described by an infinite randomness 
fixed-point, the above approximation becomes more and more accurate as the
fraction of active sites tends to zero and the SDRG is said to be 
asymptotically exact \cite{im}. 
The last effective waiting time obtained by this method $\tau_N$ 
is then in the order of the expected escape time from the
system embedded in a larger environment, or in case of an acyclic network, 
it is in the order of the mean first passage time from
site $1$ to site $N$. 
So, in case of the activated dynamics given in Eq. (\ref{activated}), 
the last exit rate must scale with $N$ as 
\be 
\ln(1/\Omega_N)\sim N^{\psi}
\label{tN}
\ee
for large $N$. 
It is easy to show that, in case of symmetric rates $w_{ij}=w_{ji}$, the
renormalization rules in Eq. (\ref{rg}) reduce to the rules of
calculating the effective resistance in the equivalent resistor network.  
This method has been used for calculating the effective resistance discussed 
in the previous section. A simplification in that case is that the sites can be eliminated in an arbitrary order.    

We have performed the above SDRG procedure numerically in finite systems and
calculated the distribution of the last effective exit rate and the average of its logarithm.   
In most of the numerical calculations, the random transition rates have been drawn
from a discrete distribution, where the rates on the edge $(ij)$ are either 
$w_{ij}=1$ and $w_{ji}=r$ with probability $1/2$ or  $w_{ij}=r$ and $w_{ji}=1$
with probability $1/2$. Here, $r$ is constant which lies in the range $0<r<1$. 
We also considered a continuous distribution, namely a  uniform one in the
domain $(0,1)$, from which all transition rates are drawn independently. 
The size of systems was $N=2^n$ with $n=6,7,\dots,13$. The number of
independent realizations was $10^6$ for $n\le 12$ and $\beta<1$, while for
larger $\beta$ and larger $N$, where the SDRG procedure is slower, 
it was less; nevertheless,
in all points it was at least a few times $10^4$.  

\begin{figure}[h]
\includegraphics[width=0.8\linewidth]{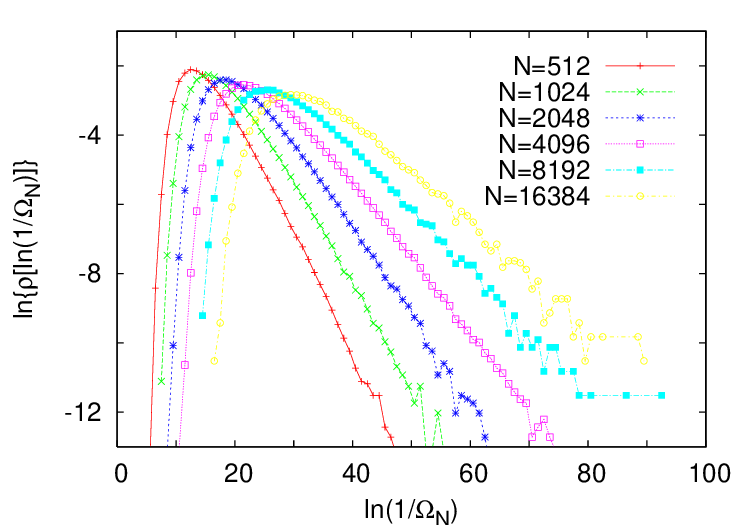}
\includegraphics[width=0.8\linewidth]{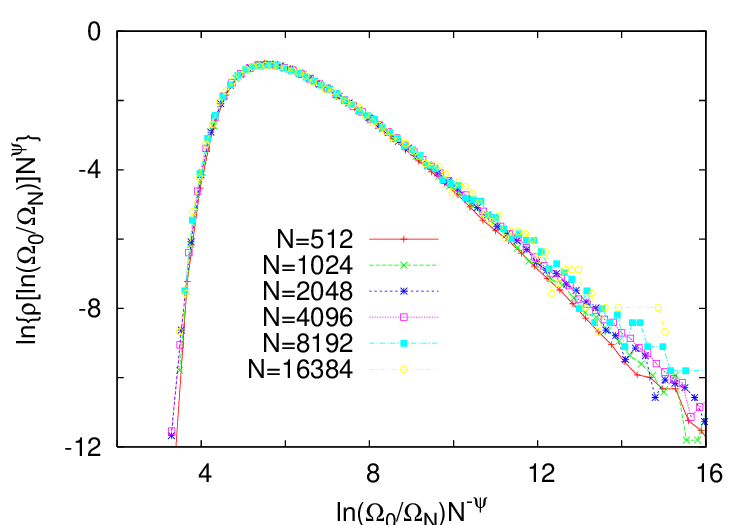}
\caption{\label{fig5} (Color online) Top: Histogram of the logarithm of the last exit rate obtained by numerical renormalization, for $s=2$, $\beta=0.7$ and for different system
sizes $N$. The rates were drawn from the discrete distribution with $r=0.1$
described in the text.
Bottom: Scaling plot of the same data with $\psi=0.19$ and $\ln\Omega_0=5.5$. 
}
\end{figure}

The distributions of $\ln(1/\Omega_N)$ for $\beta=0.7$ and for different
system sizes are shown in Fig. \ref{fig5}. As can be seen, 
they broaden with increasing $N$, which justifies the applicability of
the SDRG method a posteriori. 
The distributions are qualitatively similar for all considered values 
of $\beta$ in the range $[0,1.5]$. The probability densities have 
the scaling property 
\be
\rho[\ln(\Omega_0/\Omega_N),N]
=N^{-\psi}\tilde\rho[\ln(\Omega_0/\Omega_N)N^{\psi}],
\label{scaling}
\ee
in accordance with the assumption written in Eq. (\ref{tN}). 
Here, the  constants $\Omega_0$ and $\psi$ are found to depend on $\beta$. 
Instead of the optimal data collapse, we have estimated the barrier exponent $\psi$ systematically by the
finite-size scaling of the average $\overline{\ln(1/\Omega_N)}$.
Here, the form $\overline{\ln(\Omega_0/\Omega_N)}=cN^{\psi}$ has been
fitted to triples of data points for sizes $N$,$2N$,$4N$ and the effective
values of $\psi$ have been extrapolated to $N\to\infty$.   

\begin{figure}[h]
\includegraphics[width=0.9\linewidth]{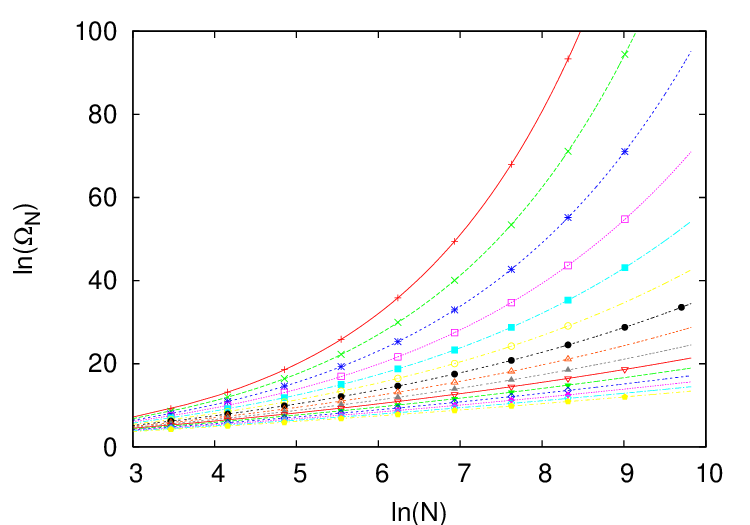}
\includegraphics[width=0.9\linewidth]{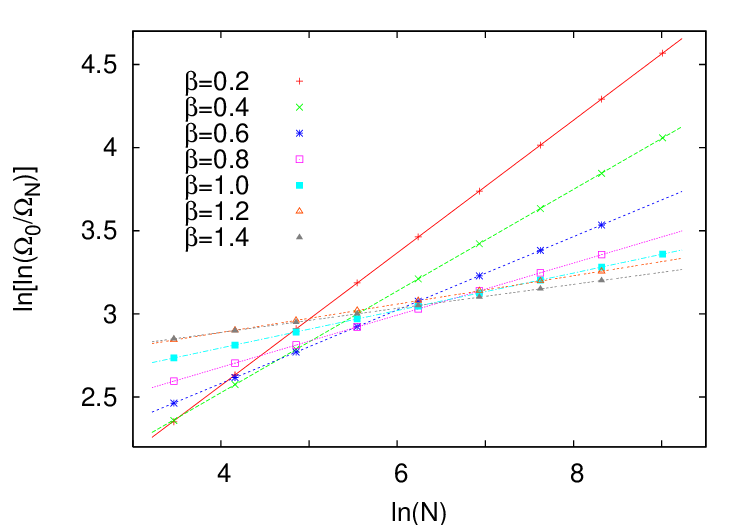}
\caption{\label{fig6} (Color online) Top: The average of the logarithm of the last exit rate
  $\overline{\ln(1/\Omega_N)}$ plotted against $\ln N$ for the same values of $\beta$ as in Fig. \ref{fig3}. 
Data has been obtained by the numerical SDRG method using 
discrete randomness with
$r=0.1$. The lines are fits of the form in Eq. (\ref{tN}) to the data. 
Bottom: The logarithm of the quantity plotted in the upper figure for certain
values of $\beta$. The slope of the straight lines is $\psi(\beta)$.   
}
\end{figure}

\begin{figure}[h]
\includegraphics[width=0.8\linewidth]{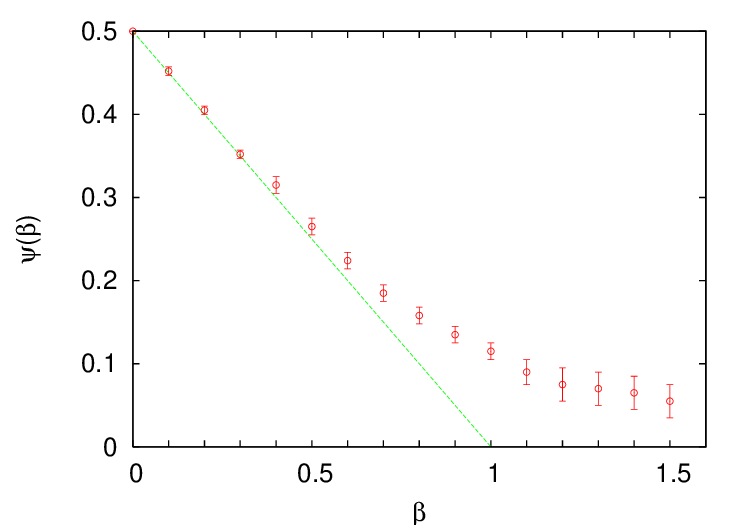}
\caption{\label{fig7} (Color online) The estimated values of $\psi(\beta)$ given in Table
  \ref{table} plotted against $\beta$. The straight line is the lower bound
  $(1-\beta)/2$.}
\end{figure}

In one-dimension it is known that the barrier exponent $\psi=1/2$ is universal,
i.e. it is independent of the distribution of transition rates. 
In order to test this property for $\beta>0$ we have studied 
the point $\beta=0.7$
with discrete randomness with $r=0.1,0.02$ and with the uniform one. 
The estimated values of $\psi$ for these three cases are in order:
$0.185(10)$, $0.189(10)$ and $0.19(1)$.
The differences are not significant, which suggests universality 
with respect the distribution of transition rates. 
So, the barrier exponent is expected to be a characteristic of the 
connectivity network. 
Next, we have determined the barrier exponents 
for different values of $\beta$, see 
Fig. \ref{fig6}, Fig. \ref{fig7} and Table \ref{table}. 
As can be seen, $\psi$ varies with $\beta$ and even for $\beta\ge 1$, 
where we do not have a lower bound on $\psi(\beta)$ at our disposal, 
the estimated values differ significantly from zero. 
\begin{table}[h]
\begin{center}
\begin{tabular}{|l|l|l|l|}
\hline   $\beta$       & $1/d_g$ & $\zeta$ & $\psi$   \\
\hline   0.1           & 0.906(2)      & 0.902(3)   &   0.452(5)      \\
\hline   0.2           & 0.825(3)      & 0.809(4)   &   0.405(5)      \\
\hline   0.3           & 0.739(4)      & 0.719(4)   &   0.352(5)      \\
\hline   0.4           & 0.667(3)      & 0.633(5)   &   0.315(10)     \\
\hline   0.5           & 0.604(3)      & 0.550(8)   &   0.265(10)     \\
\hline   0.6           & 0.552(3)      & 0.473(10)  &   0.224(10)     \\
\hline   0.7           & 0.509(3)      & 0.405(13)  &   0.185(10)     \\
\hline   0.8           & 0.475(3)      & 0.343(13)  &   0.158(10)     \\
\hline   0.9           & 0.448(3)      & 0.299(14)  &   0.135(10)     \\
\hline   1.0           & 0.426(3)      & 0.257(16)  &   0.115(10)     \\
\hline   1.1           & 0.408(4)      & 0.220(17)  &   0.09(1)       \\
\hline   1.2           & 0.393(3)      & 0.204(19)  &   0.075(20)     \\
\hline   1.3           & 0.381(3)      & 0.18(2)    &   0.07(2)       \\
\hline   1.4           & 0.371(3)      & 0.16(2)    &   0.065(20)     \\
\hline   1.5           & 0.361(3)      & 0.15(2)    &   0.055(20)     \\
\hline
\end{tabular}
\end{center}
\caption{\label{table} The estimated values of $1/d_g$, $\zeta$ and $\psi$ for
  different values of $\beta$. The estimate for $\psi$ has been obtained by
  the finite-size-scaling of $\overline{\ln(1/\Omega_N)}$ using discrete randomness with $r=0.1$.}
\end{table}

After presenting the numerical results we close this section with a heuristic
reasoning, which is based on the SDRG procedure and yields 
an upper bound on $\psi(\beta)$. 
Let us consider an acyclic network of size $N$ and find the last site, 
which the SDRG ends up with. The last effective waiting time is in the order of the
expected first passage time from site $1$ to site $N$, as we have 
already mentioned. 
Let us find now the shortest path which connects site $1$ with
site $N$ and goes through the last site obtained by the SDRG. 
Its length is $O(N^{1/d_g})$. 
Then, following the rules of the SDRG method, 
we eliminate all sites except those contained in the shortest path.
The elimination of the rest of the graph will generate new (long) edges 
between the sites of the path but the initial rates 
on the edges of the path are never decreased. 
Furthermore, it is obvious that applying the SDRG procedure to this effective
system ends up with the same site as when it was applied 
to the initial network. 
It is plausible to assume that adding long edges to a one-dimensional
system decreases the expected first passage time. 
Therefore the last effective waiting time of the system must not be longer
than that of the above one-dimensional path (without the generated
long edges). 
In the latter system the logarithm of the mean first passage time is 
$O(N^{1/(2d_g)})$, so we obtain: 
\be 
\psi(\beta)\le 1/(2d_g(\beta)). 
\ee
The numerical estimates are found to satisfy this relation for all
$\beta$, see Table \ref{table}.

\subsection{Infinite graph dimension ($1<s<2$)}

In the domain $1<s<2$, the number of edges of the graph is still $O(N)$ but the shortest path length increases with the
distance poly-logarithmically \cite{bb,biskup}. 
For even smaller decay exponent, $s<1$,
the number of edges is $O(N^{2-s})$ and the shortest path length is
bounded \cite{bb}, 
therefore we have restricted ourselves to the range $1<s<2$. 
Here, the graph dimension is formally infinite. 
According to our numerical results, the average resistance seems to saturate
with $N$, which corresponds to a vanishing resistance exponent, $\zeta=0$.   
Using Eq. (\ref{dwzeta}) it follows that the random walk with homogeneous
transition rates is ballistic here, i.e. $d_w=1$. 

\begin{figure}[h]
\includegraphics[width=0.9\linewidth]{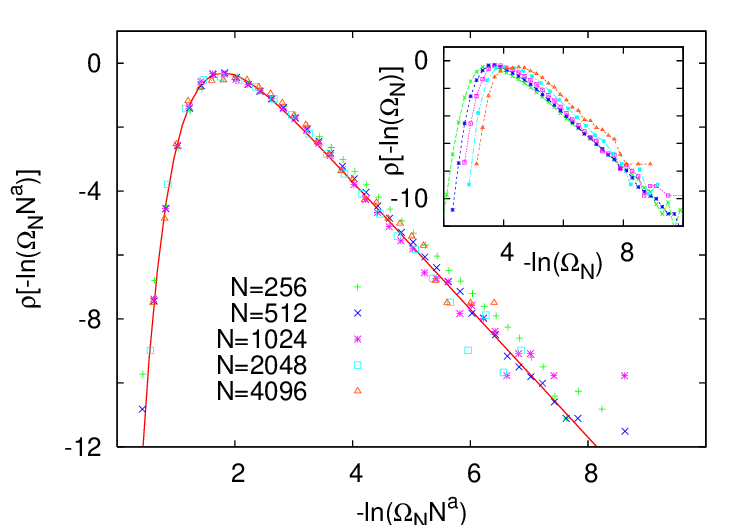}
\caption{\label{fig8} (Color online)
 Scaling plot of the distribution of the logarithm of the last exit rate obtained by numerical renormalization, for $s=1.5$, $\beta=1$ and for different system
sizes $N$. The rates were drawn from the discrete distribution with $r=0.1$. 
The scaling exponent is $a=0.3$. The solid line is the
Fr\'echet distribution given in Eq. (\ref{frechet}) with $b=2$. 
The unscaled data are shown in the inset.  
}
\end{figure}

In case of random transition rates, we have calculated the distribution of last exit rate by the SDRG method in several points. 
For $s=1.5$, $\beta=1$, the results are shown in Fig. \ref{fig8}. As can be
seen, the distributions do not broaden with increasing $N$, 
indicating 
that the system is not described by an infinite randomness fixed point.  
In this case, the waiting times neglected in the procedure may be considerable
and the last waiting time is not necessarily in the order of the first passage
time through the system but it may be only a vanishing fraction of the latter. 
The last waiting time is found to scale algebraically with $N$ as 
$\tau_N \sim N^a$, see the scaling plot in Fig. \ref{fig8}. 
It is remarkable that the data fit well 
to the Fr\'echet distribution given by the probability density 
\be 
f(\tilde\tau)=b\tilde\tau^{-1-b}e^{-\tilde\tau^{-b}},
\label{frechet}
\ee   
where $\tilde\tau=c\tau_NN^{-a}$ and $b$, $c$ are constants. 
This is well known for the random walk and other processes in one-dimensional
random environments in the driven phase \cite{jli}, where the environment can
be divided into roughly independent trapping regions with random trapping
times having the asymptotical probability density $\rho(\tau)\sim \tau^{-1-b}$.
The number of trapping regions is proportional to $N$ in one-dimension,
thus the distribution of largest waiting time follows Eq. (\ref{frechet}) 
with $a=1/b$. 
In our model with $1<s<2$, however, the latter relation seems not to be valid.
Applying the above assumption on independent trapping regions with a power law
trapping time distribution this implies that the number of effectively
independent trapping regions is not proportional to $N$ but it is 
only $O(N^{ab})$.  
We have found that the above properties of the distribution of the largest
waiting time are generally valid in the region $1<s<2$, however, the exponents
$a$ and $b$ depend on $s$ and $\beta$.

\begin{figure}[h]
\includegraphics[width=0.9\linewidth]{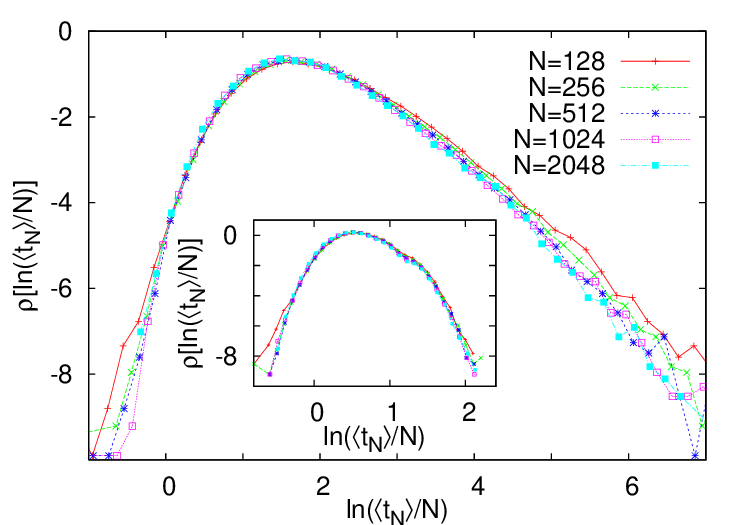}
\caption{\label{fig9} (Color online)
Scaling plot of the distribution of the logarithm of the mean first passage time $\langle t_N\rangle$ obtained
by Monte Carlo simulation in networks with $s=1.5$, $\beta=1$ for different
system sizes. For the transition rates discrete randomness has been used with
$r=0.1$. The inset shows the same quantity in the same type of networks but
with homogeneous transition rates ($r=1$).}
\end{figure}

In order to reveal the dynamics we have carried out Monte Carlo
simulations in cyclic networks and measured the first passage time from 
site $1$ to site $1+N/2$. 
For each system size, $10^5$ independent networks with random transition rates
have been generated and the mean first passage time has been calculated in
each network from $100$ measurements. 
First, we have determined the distribution of the mean first passage time in
networks with homogeneous rates. As can be seen in the inset 
of Fig. \ref{fig9}, the mean first passage time scales as 
$\langle t_N\rangle\sim N$, in accordance with $d_w=1$.
The same quantity for random transition rates is shown in the same figure. 
As can be seen, the distributions are broader than in the homogeneous case but
the motion is still ballistic, i.e. $\langle t_N\rangle\sim N$. 
Thus we conclude that although the steady state probabilities are inhomogeneous
and there are trapping regions with large waiting times in case of random
environments, the random-walk dimension is not altered compared to
the homogeneous model ($d_w=1$). So, in this sense the disorder in the transition rates is irrelevant for $s<2$.

\section{Discussion}
\label{discussion}

In this work, we have studied one-dimensional random walks
in the simultaneous presence of quenched disordered
transition rates and quenched long-range connections. 
The effects of these two components when they are present separately
are known to be antagonistic: the former induces ultra-slow activated dynamics while
the latter leads to superdiffusion. 
According to our results obtained mainly by a strong disorder renormalization
group method, the scaling exponents are 
independent of the distribution of transition rates in general 
but they do depend on the
decay exponent $s$ and the prefactor $\beta$ of the probability of 
long connections. 
For $s>2$, the long range connections are irrelevant and the dynamics follows
activated scaling with a barrier exponent $\psi=1/2$ like in one-dimensional
random environments. 
For $s<2$, where the graph dimension is formally infinite, the
disorder in the transition rates still leads to the formation 
of trapping regions with large waiting times, however,  
the dynamical exponent is not altered compared to the homogeneous environment, 
so the disorder is irrelevant in this sense.
At $s=2$, the effects of disorder and long-range connections are comparable 
and their interplay results in activated scaling with a non-trivial barrier
exponent, which varies continuously with $\beta$.    
The barrier exponent is found to decrease with $\beta$ and a positive lower
bound on $\psi(\beta)$ has been established only for $\beta<1$. 
The numerical estimates are found to be above this 
exact lower bound and the smooth decrease with $\beta$ seems to continue 
also beyond $\beta=1$ all the way to the largest value ($\beta=1.5$) studied in this work.
The numerical investigations are, unfortunately, limited to moderate $\beta$ 
since, for larger $\beta$, the number of edges is larger, as well, 
consequently the SDRG procedure becomes more and more time-consuming. 
In addition to this, the effective strength of 
disorder decreases with $\beta$, 
which would necessitate the investigation of larger and larger systems in
order to see the correct asymptotical behavior.    
So, it is still an open question whether $\psi(\beta)$
remains positive for arbitrarily large $\beta$ or it becomes zero above some
finite critical value. 

In the light of the non-trivial 
results obtained here at the point $s=2$, 
one can pose the question whether similar marginal behavior can be observed 
in the corresponding annealed model, i.e. in the simultaneous presence of quenched disordered transition rates and 
L\'evy-flight-type long-range transitions at the index $f=2$ which separates
the phases where one of the above two components is irrelevant.  
Furthermore, the marginal behavior found at $s=2$ suggests that an analogous
behavior is possible in case of more complex interacting systems with quenched
disorder and long-range connections. 

Throughout this work we have assumed that the average of the random force
acting on the particle is zero. 
On the basis of the results, we expect also the interplay between 
random-forces with non-zero average 
and long-range connections to result in non-trivial dynamics. 
The investigation of this issue is left for future work.


\acknowledgments
This paper was supported by the J\'anos Bolyai Research Scholarship of the
Hungarian Academy of Sciences and by the Hungarian National Research Fund
under grant no. OTKA K75324. 

\end{document}